\documentclass[12pt]{article}

\usepackage{scicite}

\usepackage{times}

\usepackage[utf8]{inputenc} 
\usepackage[T1]{fontenc}    
\usepackage{hyperref}       
\usepackage{url}            
\usepackage{booktabs}       
\usepackage{amsfonts}       
\usepackage{nicefrac}       
\usepackage{microtype}      
\usepackage{lipsum}
\usepackage{graphicx}
\graphicspath{ {./images/} }
\usepackage{siunitx}
\usepackage[utf8]{inputenc}
\usepackage{placeins}
\usepackage{times}
\usepackage{multirow}
\usepackage{amsmath}
\usepackage{graphicx}
\usepackage{tabularx}
\usepackage{lineno}
\usepackage[export]{adjustbox}
\usepackage{amsmath}
\usepackage{epstopdf}
\usepackage{xcolor}
\definecolor{amber}{rgb}{1.0, 0.75, 0.0}
\definecolor{brickred}{rgb}{0.7960, 0.2550, 0.3290}
\usepackage{placeins}
\usepackage{times}
\usepackage{multirow}
\usepackage{amsmath,amssymb}
\usepackage{multirow}
\usepackage{subcaption}

\newenvironment{sciabstract}{%
\begin{quote} \bf}
{\end{quote}}

\title{Unified Momentum Model for Rotor Aerodynamics Across Operating Regimes}

\author
{Jaime Liew,$^{1}$ Kirby S. Heck,$^{1}$ Michael F. Howland$^{1\ast}$\\
\\
\normalsize{$^{1}$Civil and Environmental Engineering, Massachusetts Institute of Technology,}
\\
\normalsize{$^\ast$To whom correspondence should be addressed; E-mail:  mhowland@mit.edu.}
}

\date{}

\begin{document} 


\baselineskip24pt


\maketitle


\begin{sciabstract}
Despite substantial growth in wind energy technology in recent decades, aerodynamic modeling of wind turbines relies on momentum models derived in the late 19th and early 20th centuries, which are well-known to break down under flow regimes in which wind turbines often operate. 
This gap in theoretical modeling for rotors that are misaligned with the inflow and also for high-thrust rotors has resulted in the development of numerous empirical corrections which are widely applied in textbooks, research articles, and open-source and industry design codes.
This work reports a unified momentum model to efficiently predict power production, thrust force, and wake dynamics of rotors under arbitrary inflow angles and thrust coefficients without empirical corrections. 
This unified momentum model can form a new basis for wind turbine modeling, design, and control tools from first-principles and may enable further development of innovations necessary for increased wind production and reliability to respond to 21st century climate change challenges.
\end{sciabstract}

To meet mid-century global net-zero carbon emissions targets, wind energy capacity is estimated to require between a 9- and 11-fold increase \cite{irena2022world, iea2021net}. 
Within the U.S. alone, as much as a 28-fold increase in wind power capacity is required to achieve net-zero by 2050 \cite{larson2020netzero}.
This unprecedented scale-up must be guided by modeling tools that are sufficiently accurate for design and control of wind turbines and wind farms \cite{veers2019grand, meyers2022wind}.
Despite substantial growth in wind energy technology in recent decades \cite{mai2017value}, aerodynamic modeling of wind turbine rotors relies heavily on models originally derived in the late 19th and early 20th centuries \cite{van2007lanchester}, which are well-known to break down in flow regimes that modern wind turbines often operate within \cite{glauert1926empirical, burton2011wind, sorensen1998analysis, martinez2022numerical}.
To overcome discrepancies associated with this fundamental breakdown, the predictions of wind turbine forces and power that drive contemporary design and control protocols are based on empirically fit formulas \cite{moriarty2005aerodyn, madsen2020implementation}. 
Besides numerically intensive computational fluid dynamics (CFD) simulations that have limited utility in high-throughput optimization applications, there is no existing first-principles theory that can accurately predict rotor aerodynamics across the range of thrust forces and misalignment angles commonly encountered by wind turbines. 
This work develops a unified momentum model for rotor aerodynamics that is valid across operating regimes, from low to high thrust coefficient magnitudes, including positive and negative thrust, and at arbitrary misalignment angles between inflow and rotor. 
The model overcomes limitations of classical one-dimensional momentum theory \cite{van2007lanchester} by accounting for both misalignment between the rotor and inflow and the pressure deficit in the rotor wake, as predicted by a solution to the differential Euler equations. 
The resulting aerodynamic model predicts rotor thrust, power, and wake velocities at arbitrary misalignments and thrust coefficients without empirical corrections. 
Moving beyond classical one-dimensional momentum theory and empirical corrections, this unified momentum model can form a new basis for wind turbine modeling, design, and control tools from first-principles.

One-dimensional momentum modeling, originally derived in the late 19th century by Rankine (1865) \cite{rankine1865}, W. Froude (1878) \cite{froude1878}, and R.E. Froude (1889) \cite{froude1889}, is the predominant model used in engineering analysis and design of rotors including wind turbines, propellers, helicopters, drones, and hydrokinetic turbines \cite{burton2011wind, carlton2018marine, glauert1926general, glauert1935airplane, johnson2012helicopter, amir2008modeling, pratumnopharat2011validation, ross2020experimental}. 
The theory, which is the starting point for any textbook on rotor aerodynamics across engineering applications \cite{van2007lanchester}, represents the rotor as a porous actuator disk that imparts a thrust force on the surrounding flow. 
The induced velocities generated by the rotor thrust are related to the upstream and downstream velocities via conservation of mass and momentum in one dimension that is normal to the disk.
One-dimensional momentum modeling provides reasonably accurate predictions of rotor performance at low to moderate thrust coefficients, depicted in Figure~\ref{fig:sketch} as the \textit{windmill state}. 
However, the theory is well known to break down at higher thrust coefficients (the \textit{turbulent wake state}) and for any misalignment between the inflow and the rotor, which is commonly encountered in practice \cite{stoddard1978momentum, burton2011wind}. 
In these regimes, the one-dimensional momentum theory exhibits high error for critical quantities including rotor thrust, power, wake velocities, and outlet pressure, demonstrating quantitative and qualitative deviation from measurements.
Given the one-dimensionality of the classical model, effects caused by misalignment between the rotor and the inflow are not captured.
Experiments and CFD show that thrust continues to increase with increasing induction\cite{stoddard1978momentum, wilson1974applied, glauert1926empirical}, whereas classical momentum modeling predicts the opposite trend (see Figure~\ref{fig:sketch}). 
The discrepancies arise from two main limiting assumptions: (1) one-dimensional flow perpendicular to the rotor and (2) recovery of wake pressure to the freestream value. 

The assumption of one-dimensional flow neglects all lateral velocities induced by the rotor misalignment \cite{shapiro2018modelling}.
Wind turbines continuously operate in some degree of yaw misalignment with respect to the incident wind direction due to a slowly reacting yaw controller and error or bias in wind direction measurements \cite{fleming2014field}. 
These misalignment errors are predicted to be larger for floating offshore turbines \cite{johlas2021floating}, which are anticipated to account for a large fraction of future U.S. offshore wind energy generation \cite{doeoffshorewind}.
One-dimensional momentum modeling is typically adjusted using empirical skewed wake corrections to represent the influence of rotor misalignment \cite{glauert1926general, moriarty2005aerodyn, burton2011wind}.
Textbooks state that the power production of a rotor yaw misaligned at angle $\gamma$ scales with $\cos^3(\gamma)$ \cite{burton2011wind}, while empirical observations and CFD output do not exhibit this relationship, instead showing sub-cubic behavior which varies with rotor operating conditions \cite{dahlberg2005research, liew2020analytical, howland2020influencefield}.
In tandem, given the increasing proliferation of wind energy and the densification of wind farms that leads to negative aerodynamic wake interactions between neighboring turbines \cite{barthelmie2007modelling, barthelmie2010evaluation}, research has focused on methods to collectively operate turbines within a farm by controlling the wind flow to maximize farm power production \cite{meyers2022wind}. 
The primary wind farm flow control methodology, termed wake steering, entails intentionally yaw misaligning wind turbines with respect to the freestream wind \cite{fleming2019initial, howland2022collective}, a misalignment that results in an explicit break down of the one-dimensional momentum modeling used to predict the power, loads, and wake velocities associated with the yawed turbine.

The wake pressure recovery assumption is violated at higher thrust coefficients where the static pressure in the wake remains lower than freestream pressure. 
This effect, known as base suction \cite{roshko1955wake}, corresponds to additional thrust not captured by current theoretical models.
Specifically, classical momentum modeling is widely understood and accepted to break down at a value of the induction factor ($a=1-u_d/u_\infty$, where $u_d$ is the velocity at the rotor disk and $u_\infty$ is the freestream wind speed) that is only $10\%$ higher than the optimal value of $a=1/3$ \cite{betz1920, van2007lanchester} predicted by Betz (1920) \cite{betz1920,burton2011wind, wilson1974applied}, based on classical one-dimensional momentum modeling.
There is no comprehensive theoretical or analytical model rooted in first-principles that can adequately capture the effects caused by rotor misalignment and wake pressure.
This gap in theoretical modeling has prompted the development of numerous empirical corrections, which have found widespread application in textbooks \cite{burton2011wind, eggleston1987wind}, research articles \cite{pratumnopharat2011validation, madsen2020implementation, buhl2005new}, and open-source, as well as industry, design codes \cite{moriarty2005aerodyn, bay2020floris, jonkman2005fast}.
Examples include the empirical Glauert correction to model the high thrust \textit{turbulent wake state} \cite{glauert1926empirical, pratumnopharat2011validation, madsen2020implementation, buhl2005new, burton2011wind, sorensen1998analysis, martinez2022numerical, moriarty2005aerodyn}, skewed wake corrections \cite{glauert1926general, moriarty2005aerodyn, burton2011wind}, and an empirically tuned power-yaw formula to model rotor misalignment \cite{liew2020analytical, gebraad2016wind, fleming2019initial}.

This work extends first-principles understanding and develops a new analytical aerodynamic model for rotors operating in arbitrary thrust and misalignment conditions. 
To address long-standing discrepancies between classical momentum modeling and empirical observations, the actuator disk model is re-examined from first principles. 
Using conservation of mass, momentum, and energy, the limiting assumptions of the classical theory are eliminated by modeling the pressure deficit in the rotor wake, using a solution to the differential Euler equations, and by accounting for arbitrary rotor misalignment with a lifting line model. 
This results in a set of five coupled equations that simultaneously govern rotor induction, thrust, wake velocities, wake pressure, and power production. 
The unified equations predict the empirically observed monotonic increase in thrust coefficient with increasing induction factor, offering a first-principles solution to a long-standing limitation of classical one-dimensional momentum modeling that predicts a trend in the opposite direction. 
The equations further predict the quantitative influence of rotor misalignment on wake velocities, thrust, and power at arbitrary, joint misalignments and thrust coefficients.
This study provides new theoretical insight and a simple, computationally-efficient model for complex turbine aerodynamics that have previously relied on empirical corrections or necessitated expensive CFD to resolve \cite{sorensen1998analysis, martinez2022numerical}.

\subsection*{Unified momentum model}

In this study, we derive an analytical relationship between the induction, streamwise and spanwise wake velocities, and the wake pressure through integral analysis of a control volume enclosing the actuator disk. 
This unified momentum theory addresses the two limiting assumptions of one-dimensionality and pressure recovery to freestream from classical momentum modeling.
The induction is modeled using the Bernoulli equation, with the pressure drop at the actuator disk predicted by the thrust force. 
The streamwise wake velocity is modeled using momentum conservation in the streamwise direction, along with mass conservation in the control volume and the streamtube enclosing the actuator disk.
Both the induction and the streamwise wake velocity depend on each other, and depend on the pressure deficit in the wake.
Given an arbitrary misalignment angle between the actuator disk and the incident flow, the flow also contains a lateral component and is not one-dimensional. 
We model the lateral wake velocity using a lifting line model \cite{milne1973theoretical, shapiro2018modelling, heck2023modelling}, that relates the lateral wake velocity to the lift component of the rotor-normal thrust force.

The remaining unknown to close the system of equations is the pressure in the wake, which classical one-dimensional momentum modeling assumes to equal freestream pressure \cite{burton2011wind}.
As the induced velocity is increased with increasing thrust coefficients, flow separation and a large pressure drop can occur\cite{roshko1955wake, steiros2018drag}, which prevents the wake pressure from recovering to its freestream value, and correspondingly increases the thrust \cite{burton2011wind}.
To address this, the wake pressure in this study is modeled using a solution to the steady, two-dimensional Euler equations with an actuator disk turbine body forcing, recently proposed by Madsen (2023) \cite{madsen2023linear}, based on the analytical method of Von K{\'a}rm{\'a}n and Burgers (1935) \cite{durand1935aerodynamic}.
The solution to the pressure Poisson equation is decomposed into contributions from the linear actuator disk body forces, and the nonlinear advective terms.
To complete this set of equations, a model for the near-wake length is needed to determine the pressure in the wake at the outlet of the inviscid near-wake region.
Bastankhah \& Port{\'e}-Agel (2016) \cite{bastankhah2016experimental} used the shear layer model of Lee \& Chu (2003) \cite{lee2003turbulent} and one-dimensional momentum modeling to predict the near-wake length for low thrust and induction rotors.
To predict the near-wake length, we generalize this approach to be consistent with high-thrust and induction and yaw misaligned regimes.
The shear layer model introduces a single parameter, $\beta$, that relates the shear layer growth to the characteristic velocity shear between the wake and freestream velocity \cite{bastankhah2016experimental}.
This parameter has a well-accepted range in the literature \cite{lee2003turbulent}. 
Here, the parameter is estimated using data from CFD of an actuator disk and we find that it is within the standard range.
Further, the quantitative influence of the parameter uncertainty is assessed in the results below, and in {\bf Supplementary Information}.
The final system of equations, given in {\bf Methods}, provides an analytical relationship between the induction, streamwise velocity, lateral velocity, and wake pressure for a rotor with arbitrary yaw misalignments and thrust coefficients.
The corresponding power production for a wind turbine or hydrokinetic rotor is computed using the predicted induction factor.

\subsection*{Results}

\begin{figure}
    \centering\includegraphics[width=0.8\linewidth]{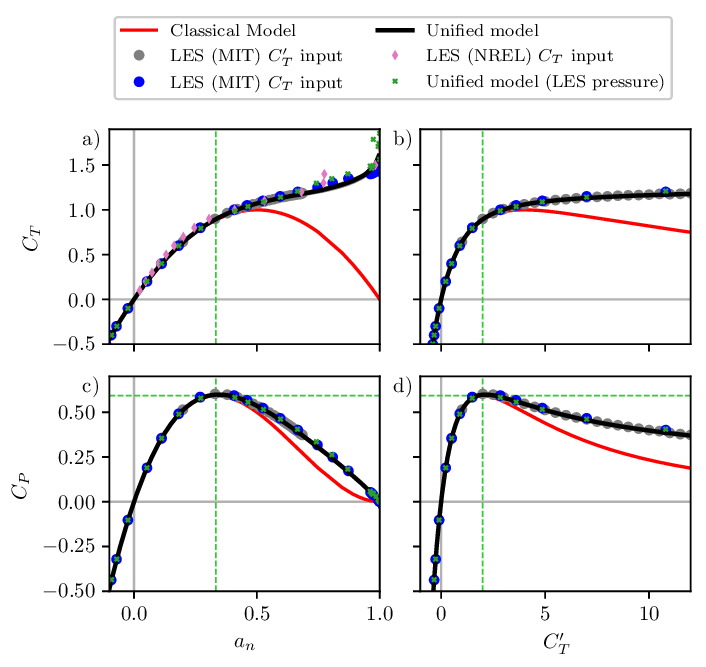}
    \caption{Coefficient of thrust, $C_T$, as a function of (a) rotor-normal induction factor, $a_n$, and (b) the local thrust coefficient $C_T^\prime$. Coefficient of power, $C_p$, as a function of (c) $a_n$, and (d) $C_T^\prime$. The variables estimated by the presented unified model ({\bf Methods}), are shown. 
    The shaded region corresponds to $\pm 10\%$ uncertainty in parameter $\beta$.
    Results from large eddy simulation (LES) from MIT and NREL and classical one-dimensional momentum modeling are shown as a reference. 
    {\it A priori} model results are also shown for reference, where the LES measured pressure deficit $(p_1-p_4)/\rho$ is provided as an input to the model to facilitate prediction of $C_T$ and $C_P$.
    The Betz limit of $a_n=1/3$ ($C_T^\prime=2$) and $C_P=16/27$ is also shown by the dashed green line.
	}
	\label{fig:disk}
\end{figure}

Predictions from the model proposed here are assessed through {\it a priori} and {\it a posteriori} analysis, compared to high-fidelity large eddy simulation (LES) CFD of an actuator disk modeled rotor.
The numerical implementation of the LES is described in {\bf Methods}.
To demonstrate the robustness of the numerical method, LES results are shown using two standard approaches for modeling the thrust force (see Calaf {\it et al.} \cite{calaf2010large}), depending on input of either the thrust coefficient $C_T$ and a thrust force dependent on the freestream velocity $\vec{u}_\infty$, or input of a modified thrust coefficient $C_T^\prime$ and a thrust force dependent on the velocity at the rotor $\vec{u}_d$.
Figure~\ref{fig:disk}(a) illustrates the thrust coefficient $C_T$ as a function of the rotor-normal induction factor $a_n$ ($a_n = 1 - \vec{u}_d \cdot \hat{n}/ \vec{u}_\infty \cdot \hat{n}$), which is the generalization of the classical one-dimensional induction factor.
In addition to the present LES results, LES results from Mart{\'i}nez-Tossas {\it et al.} (2022, NREL) \cite{martinez2022numerical} are shown for reference.
To demonstrate the validity of the model form, {\it a priori} model predictions are shown, where the wake pressure in LES is measured and provided to the model. 
The fully-predictive {\it a posteriori} model output, where the pressure is modeled based on the methodology described previously, are also shown.
The {\it a priori} model output exhibits remarkable agreement with LES in terms of predicting induction, wake velocities and power production, validating the model-form proposed here.
The fully-predictive model, which requires no pressure input from LES, exhibits low error across the full range of rotor-normal induction factors realized in LES, showing substantial qualitative and quantitative predictive accuracy improvements compared to classical one-dimensional momentum modeling.
The clear and notable deficiency of classical momentum modeling to predict the qualitative and quantitative response of the thrust coefficient $C_T$ for induction factors of $a_n \gtrsim 1/3$ has resulted in numerous empirical formulas to be proposed in the literature \cite{moriarty2005aerodyn,pratumnopharat2011validation, madsen2020implementation, buhl2005new}.
In this study, through first-principles modeling of the wake pressure, substantial increases in predictive accuracy are achieved without empirical formulas.
Similar observations can be made through analysis of the coefficient of power $C_P$ depending on the rotor-normal induction factor $a_n$ is shown in Figure~\ref{fig:disk}(c).

Interestingly, the proposed model predicts the maximum value of $C_P$ as $0.5984$, occurring at an induction factor of $0.345$, which is approximately $1.0\%$ and $3.5\%$ higher than the classical Betz limit \cite{betz1920, van2007lanchester} predictions for the coefficient of power ($16/27$) and power-maximizing induction factor ($1/3$), respectively. 
Since the Betz limit is derived using classical one-dimensional momentum modeling, assuming that the wake pressure recovers to freestream, it inherently dictates that the turbine cannot extract energy from pressure. In contrast, the model presented here does account for the persistent pressure drop in the far wake, resulting in a marginal increase in energy extraction.
While these are modest departures from the classical Betz limit, this result elucidates that neglecting the energetic contributions of the wake pressure deficit will incorrectly model energy conservation. 
This result is further highlighted by the consistent underprediction of the coefficient of power $C_P$ of one-dimensional momentum modeling for induction factors of $a_n \gtrsim 1/3$.
The consistent one-dimensional momentum modeling error in the coefficient of power for $a_n \gtrsim 1/3$ is alleviated by the pressure model proposed here.
In Figure~\ref{fig:disk}(b,d), the coefficients of thrust and power are shown as a function of the modified thrust coefficient $C_T^\prime$ \cite{calaf2010large} for $0<C_T^\prime<12$, which is a common alternative method to model actuator disks in CFD.
As is evident in Figure~\ref{fig:disk}(a), the marginal increase in $C_T$ is reduced for higher values of $C_T^\prime$, such that $C_T^\prime=12$ results in an induction factor of only $a_n\approx 0.7$. 
This results from the definition of the thrust described previously, where the thrust force depends on the local thrust coefficient $C_T^\prime$ and the disk velocity $\vec{u}_d$.
As the local thrust coefficient $C_T^\prime$ is increased, the induction also increases. 
Correspondingly, the disk velocity $\vec{u}_d$ is lowered, which in turn lowers the thrust force. 
The counteracting effects of increasing $C_T^\prime$ and decreasing disk velocity limits the growth in the thrust coefficient $C_T$, which depends on these variables jointly.
Since rotors produce thrust force and power based on the velocity at the rotor ($\vec{u}_d$), the local thrust coefficient model is a more physically-consistent representation of a wind turbine rotor \cite{heck2023modelling, calaf2010large}.

\begin{figure}
    \centering\includegraphics[width=0.8\linewidth]{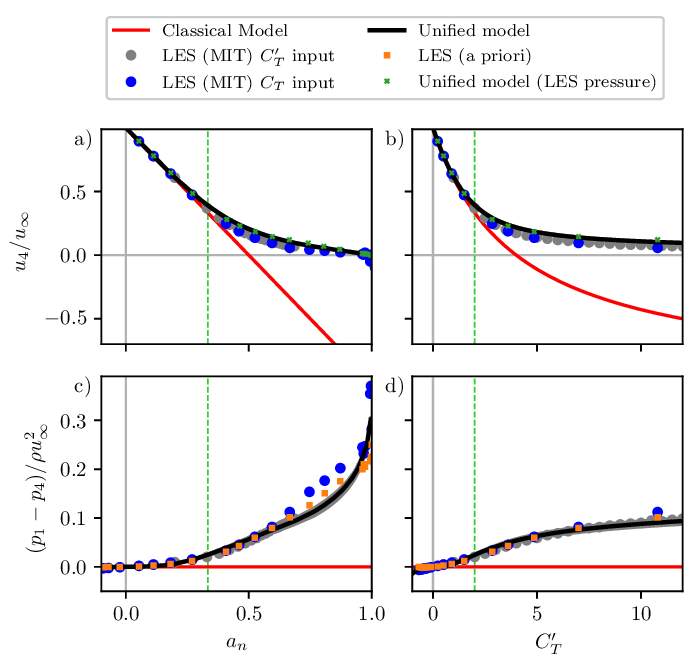}
\caption{Wake streamwise velocity $u_4$ as a function of (a) rotor-normal induction factor, $a_n$, and (b) local thrust coefficient $C_T^\prime$. Density-normalized wake pressure deficit $(p_1-p_4)/\rho$ as a function of c) $a_n$ and d) $C_T^\prime$. The variables estimated by the presented unified momentum model model equations ({\bf Methods}), are shown.  
    The shaded region corresponds to $\pm 10\%$ uncertainty in $\beta$.
    Results from LES and classical one-dimensional momentum modeling are shown as a reference.
    Two additional {\it a priori} model results are shown. 
    For the wake streamwise velocity $u_4$ (a, b), {\it a priori} model results are shown where the LES measured pressure deficit $(p_1-p_4)/(\rho u_\infty^2)$ is provided as an input to the model to facilitate prediction of wake velocity $u_4$.
    For the wake pressure deficit $(p_1-p_4)/(\rho u_\infty^2)$ (c, d), the pressure deficits that are implied from energy conservation, with input of LES measured $a_n$ and $u_4$, are shown.
    The Betz limit of $a_n=1/3$ and $C_T^\prime=2$ are also shown by the dashed green line.
	}
	\label{fig:wake}
\end{figure}

Since the modeling framework proposed here represents the relationship between thrust, power, and wake variables from first-principles, the predicted wake velocities conserve mass, momentum, and energy across arbitrary thrust coefficients and misalignment angles.
The streamwise wake velocity and the density-normalized wake pressure deficit are shown in Figure~\ref{fig:wake}(a,c), respectively, as a function of the rotor-normal induction $a_n$, and in Figure~\ref{fig:wake}(b,d) as a function of the modified thrust coefficient $C_T^\prime$.
Above $a_n=0.5$ and $C_T^\prime=4$, classical momentum modeling predicts that the wake velocity is negative, which differs from the LES output.
Instead, both the LES and the model predictions ({\it a priori} and {\it a posteriori}) asymptote to zero wake velocity for $a_n=1$ ($C_T^\prime\rightarrow\infty$). 
The density-normalized wake pressure deficit (Figure~\ref{fig:wake}(b)) from LES is first compared to the wake pressure that is implied by energy conservation in the control volume (further described in {\bf Supplementary Information}). 
To estimate the wake pressure implied by energy conservation, we solve for the wake pressure deficit using the model equation for the induction factor $a_n$, and we input the LES measured values for $a_n$ and the wake velocities.
As shown in Figure~\ref{fig:wake}(b), there is excellent agreement between the LES measured wake pressure and the wake pressure implied by energy conservation for $a_n<0.7$, further validating the model form.
As $a_n$ approaches unity, the accuracy of the Bernoulli equation in the wake is reduced, which lowers the quantitative accuracy of the pressure prediction in this region, but this has a limited quantitative impact on the primary variables of interest ($C_T$, $C_P$, and wake velocities shown in Figure~\ref{fig:disk} and Figure~\ref{fig:wake}(a)).
Finally, we show the fully predictive wake pressure results from the differential Euler equations and near-wake length closure proposed here.
The predictive pressure model exhibits qualitative and quantitative agreement, with some increasing predictive error at values of induction above $a_n>0.7$ from the Bernoulli equation as described previously.
As the induction $a_n$ approaches unity, there is a nonlinear increase in the wake pressure deficit, with the limiting state of $a_n=1$ resulting in flow separation \cite{roshko1955wake, steiros2018drag}.
At and above induction of unity ($a_n\geq1$), the wake flow is separated as indicated in Figure~\ref{fig:sketch}, resulting in the bluff body wake of a flat circular plate. In such cases, the presented model is no longer valid.

A parallel modeling approach to the integral analysis presented here represents the porous disk with a distribution of equal magnitude sources in potential flow \cite{taylor1944air, koo1973fluid, o2006source}.
However, these approaches have also in large part neglected the wake pressure deficit, similar to classical one-dimensional momentum modeling.
Steiros \& Hultmark (2018) \cite{steiros2018drag} extended the source modeling approach by providing a more detailed representation of the wake, using momentum conservation and the Bernoulli equation, and by including a wake pressure term.
To close the derived system of equations analytically without an additional pressure model equation, a simple wake factor was introduced to model the streamwise wake velocity.
The one-dimensional model proposed by Steiros \& Hultmark (2018) \cite{steiros2018drag} exhibited excellent agreement compared to experimental measurements of the coefficient of drag (thrust) of porous plates immersed a water tank.
While the thrust predictions of this model exhibit substantial improvements for high values of induction compared to classical one-dimensional momentum modeling \cite{steiros2018drag}, it is notable that the wake factor-based wake velocity model differs from classical one-dimensional momentum model predictions and CFD data, even for low induction values (below the heavily loaded limit of $a_n\approx 0.37$ \cite{wilson1974applied}).
Detailed comparisons between the model proposed here and the Steiros \& Hultmark (2018) \cite{steiros2018drag} model are made in the {\bf Supplementary Information}.

\begin{figure}
    \centering\includegraphics[width=1\linewidth]{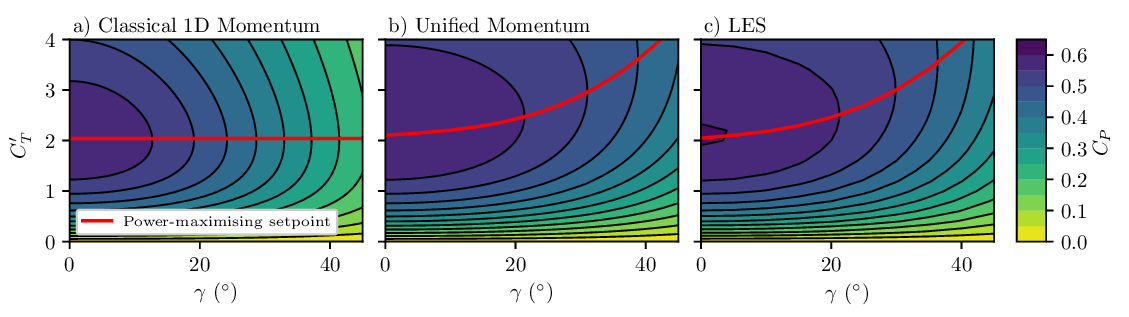}
	\caption{Coefficient of power $C_P$ as a function of the yaw misalignment $\gamma$ and the local thrust coefficient $C_T^\prime$ for (a) classical one-dimensional momentum modeling and empirically modeling the effect of yaw misalignment as $P(\gamma)=P(\gamma=0)\cdot \cos^3(\gamma)$ which is a common model \cite{burton2011wind} given the previous lack of a first-principles approach, (b) the presented unified momentum theory ({\bf Methods}), and (c) LES. The power-maximizing $C_T'$ set points as a function of $\gamma$ are indicated in red.
	}
	\label{fig:Cp_contour}
\end{figure}

We further evaluate the model predictions for yaw misaligned rotors.
Accurately predicting the power output of a yaw-misaligned wind turbine is crucial, given that turbines typically operate in yaw \cite{fleming2014field}, and for the effective flow control in wind farms through wake steering \cite{howland2022collective, meyers2022wind}.
Recently, Heck {\it et al.} (2023) \cite{heck2023modelling} demonstrated that the deviation of yawed rotor power production from $\cos^3(\gamma)$ results from the impact of the yaw misalignment on the induction factor $a_n$, but the proposed model had increasing error with increasing thrust coefficients because it assumed that the wake pressure recovers to freestream pressure. The presented unified momentum model extends the framework introduced by Heck et al. (2023), ensuring its validity at high thrust coefficients by relaxing the pressure recovery assumption.
In Figure~\ref{fig:Cp_contour}, the coefficient of power $C_P$ is shown for yaw misalignments between $0<\gamma<60^\circ$ and local thrust coefficients between $0<C_T^\prime<4$ for the proposed model and for LES, along with a baseline model from classical one-dimensional momentum modeling.
The model proposed here exhibits high levels of qualitative and quantitative accuracy compared to LES.
Specifically, the model demonstrates an $84\%$ and an $21\%$ reduction in the mean absolute error across the thrust coefficient and yaw misalignment values considered here compared to classical one-dimensional momentum modeling (with $P(\gamma)=P(\gamma=0)\cdot \cos^3(\gamma)$) and to the model that neglects the wake pressure deficit \cite{heck2023modelling}, respectively.

\subsection*{Discussion}

The simple, first-principles analytical relationship between the induction, thrust, power, wake velocities, and wake pressure proposed here is ideally suited for implementation in widely used blade-element momentum (BEM) modeling frameworks for wind and hydrokinetic turbines and propellers, such as OpenFAST \cite{jonkman2005fast} and HAWC2 \cite{larsen20072}, as well as in wake models used for wind farm design and control, such as the FLORIS model \cite{bay2020floris} and PyWake \cite{pedersen2019dtuwindenergy}.
The predictive model developed here, which has a runtime on the order of milliseconds on a standard desktop computer, has lower predictive error than classical one-dimensional momentum modeling for all thrust coefficients and misalignment angles considered, enabling rotor modeling for arbitrary thrust and yaw conditions without empirical corrections for the first time. 
The relationship improves our physical understanding of rotors operating across all thrust regimes, including a modification to the Betz limit \cite{van2007lanchester, betz1920} that is enabled by momentum and kinetic energy extraction from the pressure.
Future work is encouraged to enhance the understanding of the shear layer growth parameter, $\beta$. 
Although this parameter exhibits a weak sensitivity on rotor quantities such as thrust and power, it exerts a more significant impact on wake variables, particularly governing the balance between wake pressure and wake velocity (see {\bf Supplementary Information}). 
Further investigation is recommended to explore the universality of this parameter in comparison to high Reynolds number experimental measurements.

Wind turbines are increasing in rotor diameter, hub-height, and design and control complexity, pushing the limits of applicability of existing modeling tools, and motivating the scientific community to address grand scientific challenges \cite{veers2019grand}, focused on the design \cite{veers2023grand} and control \cite{meyers2022wind} of modern utility-scale wind turbines and wind farms.
The proposed unified momentum model is a first-principles-based approach that unifies rotor thrust, yaw, and induction with the outlet velocity and pressure of flow through an actuator disk. It addresses major theoretical limitations in modeling rotor performance under high thrust and yaw misalignment. These complex operating conditions have historically relied on empirical models due to gaps in underlying theory. Derived under assumptions of uniform, zero-turbulence inflow, it provides an ideal starting point for extensions to more complex operational regimes such as non-uniform inflow and rotational effects, rather than relying on empirical models that are unlikely to extrapolate out-of-sample. 
Future work should extend the analysis and model to consider unsteadiness, such as from floating motion for an offshore turbine, wind speed and direction shear, and turbulence.

\clearpage
\FloatBarrier
\section*{Figures}

\begin{figure}[!htb]
    \centering
    \makebox[\linewidth][c]{%
        \includegraphics[width=1.2\linewidth]{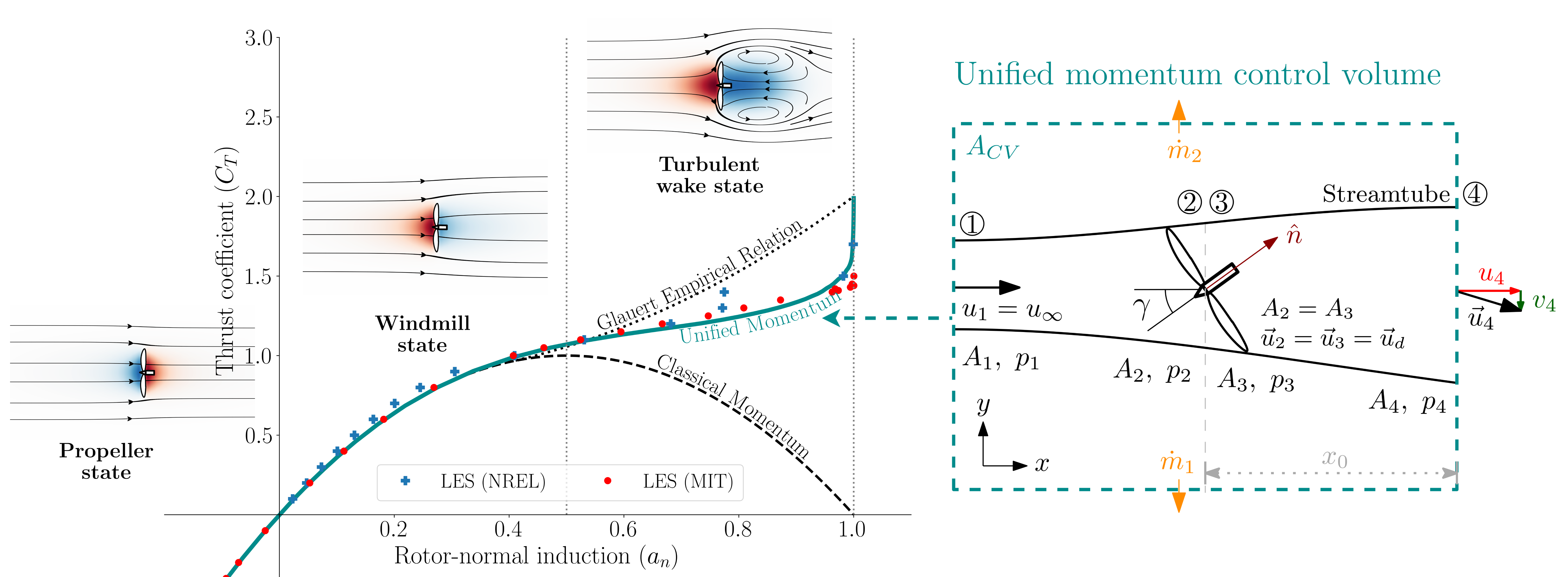}%
    }
    \caption{(Left) Schematic illustrating the rotor thrust coefficient variations with rotor-normal induction across various operational scenarios (propeller state, windmill state, and turbulent wake state \cite{stoddard1978momentum}) for a yaw aligned actuator disk. 
    Model predictions are shown using classical one-dimensional momentum modeling \cite{rankine1865, froude1878, froude1889}, Glauert's empirical relation \cite{glauert1926empirical}, and the unified momentum model introduced in this study. 
    (Right) Schematic representation of the control volume enclosing the porous actuator disk that is used to derived the unified momentum model developed in this study.}
    \label{fig:sketch}
\end{figure}

\FloatBarrier



\FloatBarrier

\section*{Acknowledgments}
K.S.H. and M.F.H. acknowledge funding from the National Science Foundation (Fluid Dynamics program, grant number FD-2226053, Program Manager: Dr. Ronald D. Joslin).
J.L. acknowledges support from Siemens Gamesa Renewable Energy.
All simulations were performed on Stampede2 supercomputer under the NSF ACCESS project ATM170028.

\section*{Data and code availability}
A reference implementation of the unified momentum model, as well as all figure-generating code is available as an open-source Python library at \url{https://github.com/Howland-Lab/Unified-Momentum-Model}\cite{liew_2024_10524067}. 
The nonlinear centerline pressure, $p^{NL}$ is also tabulated in this library. 
The large eddy simulation software source code, Pad{\'e}Ops\cite{ghate2018padeops}, is available at \url{https://github.com/Howland-Lab/PadeOps}.

\section*{Methods}

\subsection*{Unified momentum model}
This study employs integral analysis of the control volume enclosing the actuator disk (Figure~\ref{fig:sketch}) to derive a system of analytical equations describing the relationships between rotor-normal induction, thrust force, power production, wake pressure, and wake outlet velocities.
The derivation is described in detail in the model derivation section of the {\bf Supplementary Information}.
The analytical equations consider both yaw aligned and yaw misaligned (porous) actuator disks. 
The inputs are the modified thrust coefficient $C_T'$ and yaw misalignment angle $\gamma$.
The equations then solve for the rotor-normal induction $a_n$, streamwise wake velocity $u_4$, lateral wake velocity $v_4$, near-wake length, $x_0$, and the pressure difference $(p_4-p_1)$ as outputs. 
The final form of the equations are:
\begin{align}
\label{eq:final_1}
a_n &= 1 - \sqrt{\frac{u_\infty^2 - u_4^2 - v_4^2}{C_T^\prime \cos^2(\gamma) u_\infty^2}-\frac{(p_4 - p_1)}{\frac{1}{2} \rho C_T^\prime \cos^2(\gamma) u_\infty^2}}\\ \label{eq:final_2}
u_4 &= -\frac{1}{4} C_T^\prime (1-a_n) \cos^2(\gamma)u_\infty + \frac{u_\infty}{2} + \frac{1}{2}\sqrt{\left(\frac{1}{2} C_T^\prime (1-a_n) \cos^2(\gamma)u_\infty - u_\infty \right)^2 - \frac{4(p_4 - p_1)}{\rho}}\\ \label{eq:final_3}
v_4 &= -\frac{1}{4} C_T^\prime (1-a_n)^2 \sin(\gamma) \cos^2(\gamma) u_\infty\\ \label{eq:final_4}
\frac{x_0}{D} &=\frac{\cos(\gamma)}{2\beta}\cfrac{u_\infty + u_4}{|u_\infty - u_4|}\sqrt{\frac{(1-a_n)\cos(\gamma)u_\infty}{u_\infty + u_4}} \\
\label{eq:final_5}
p_4 - p_1 &= -\frac{1}{2\pi} \rho C_T^\prime (1-a_n)^2 \cos^2(\gamma) u_\infty^2 \arctan\left[\frac{1}{2}\frac{D}{x_0}\right] + p^{NL}(C_T', \gamma, a_n, x_0),
\end{align}
where the freestream incident wind speed is $u_\infty$, the fluid density is $\rho$, the actuator disk diameter is $D$, and the unknown wake spreading rate is $\beta=0.1403$ as outlined in the \textbf{Supplementary Information}. 
The pressure equation (Eq.~\eqref{eq:final_5}) contains two terms.
The first term is the pressure contribution from the actuator disk forcing, and the second term ($p^{NL}$) is a nonlinear term that results from the advection.
The variable $p^{NL}$ is the nonlinear pressure contribution to the outlet wake pressure, which is described in detail in the model derivation in \textbf{Supplementary Information}.

In some applications, it can be preferable to use the thrust coefficient $C_T$ as an input, rather than the local thrust coefficient $C_T^\prime$.
For example, in blade-element modelling of a rotor, or if a utility-scale wind turbine thrust curve is available as a function of the freestream wind speed $u_\infty$, using $C_T$ as the input can be more convenient.
When solving using $C_T$ as an input instead of $C_T'$, a sixth equation is added:
\begin{equation}
\label{eq:final_6}
    C_T' = \frac{C_T}{(1-a_n)^2\cos^2(\gamma)},
\end{equation}
where the thrust coefficient is $C_T = 2 {\|F_T\|} / (\rho A_d u_\infty^2),$ where ${\|F_T\|}$ is the magnitude of the thrust force and $A_d = \pi D^2/4$ is the rotor area.
The system of equations can be written with $C_T$ as the input variable
\begin{align}
\label{eq:pressure_1}
a_n &= 1 + \frac{1}{2} \frac{C_T u_\infty^2}{u_\infty (u_4 - u_\infty) + \frac{u_\infty}{\rho u_4} (p_4-p_1)} \\
\label{eq:pressure_2}
u_4 &= u_\infty \sqrt{1 - \frac{1}{16} C_T^2 \sin^2(\gamma) - C_T - \frac{2}{\rho u_\infty^2} (p_4-p_1)} \\
\label{eq:pressure_3}
v_4 &= -\frac{1}{4} C_T \sin(\gamma) u_\infty \\
\label{eq:pressure_4}
\frac{x_0}{D} &=\frac{\cos(\gamma)}{2\beta}\cfrac{u_\infty + u_4}{|u_\infty - u_4|}\sqrt{\frac{(1-a_n)\cos(\gamma)u_\infty}{u_\infty + u_4}} \\
\label{eq:pressure_5}
p_4 - p_1 &= -\frac{1}{2\pi} \rho C_T u_\infty^2 \arctan\left[\frac{1}{2}\frac{D}{x_0}\right] + p^{NL}(C_T, \gamma, a_n, x_0).
\end{align}

\subsection*{Large eddy simulation numerical setup}
Large eddy simulations are performed using Pad{\'e}Ops \cite{ghate2017subfilter, howland2020influence}, an open-source incompressible flow code \cite{ghate2018padeops}.
The horizontal directions use Fourier collocation, and a sixth-order staggered compact finite difference scheme is used in the vertical direction \cite{nagarajan2003robust}.
A fourth-order strong stability preserving (SSP) variant of Runge-Kutta scheme is used for time advancement \cite{gottlieb2011strong} and the sigma subfilter scale model is used \cite{nicoud2011using}.
Simulations are performed with uniform inflow with zero freestream turbulence, consistent with the derivation of classical momentum modeling.
The boundary conditions are periodic in the $x$ and $y$ directions with a fringe region \cite{nordstrom1999fringe} used in the $x$-direction to remove the wake from recirculating.
Periodic boundary conditions are used in the lateral $y$-direction.
The simulations are performed with a domain size of $L_x=25D$ in length and cross-sectional size $L_y = 20D, L_z=10D$ with $256\times512\times256$ grid points and with the turbine $5D$ downwind of the inlet. 

The porous disk is modeled using an actuator disk model (ADM), that imparts a thrust force that depends on the modified thrust coefficient $C_T^\prime$ and the disk velocity $\vec{u}_d \cdot \hat{n}$ \cite{calaf2010large}
\begin{equation}
    \vec{F}_T = -\frac{1}{2} \rho C_T^{\prime} A_d (\vec{u}_{d} \cdot \hat{n})^2 \hat{n},
\label{eq:ft_full_les_code}
\end{equation}
where $\rho$ is the air density, $A_d=\pi D^2/4$ is the area of the disk where $D$ is the diameter, and $\hat{n}$ is the unit normal vector perpendicular to the disk. 
Note that this differs from the simplified model of thrust force from an actuator disk that depends on the freestream rotor-normal wind speed, $\vec{u}_\infty$, $\vec{F}_{T,\mathrm{ideal}} = -\frac{1}{2} \rho C_T A_d \|\vec{u}_{\infty}\|^2 \hat{n}$, where $\vec{u}_{\infty}=u_\infty \hat{\imath}+ 0 \hat{\jmath}$ is the freestream wind velocity vector and $C_T=-\|\vec{F}_{T,\mathrm{ideal}}\|/\frac{1}{2} \rho A_d \|\vec{u}_{\infty}\|^2$.
Porous disks produce thrust based on the wind velocity at the rotor, which has been modified by induction.
The coefficient of thrust $C_T$ is an empirical quantity, that depends on the magnitude of the induction, and it needs to be measured or predicted using a model.
It is preferable for both analytical and numerical modeling, and more physically intuitive, to model the thrust force based on the velocity that is accessible to the porous disk, which is the disk velocity $\vec{u}_{d}$, and thus $C_T^\prime$ is the input thrust coefficient.
For a yaw aligned turbine, $C_T^\prime = C_T/(1-a(\gamma=0))^2$ \cite{calaf2010large}. 

The disk-normal, disk-averaged induction factor $a_n$ for a disk with yaw misalignment angle $\gamma$ is defined as
\begin{equation}
    a_n = 1 - \frac{ \vec{u}_d \cdot \hat{n} }{ u_{\infty} \cos(\gamma)}. 
    \label{eq:an_code}
\end{equation}
In the yaw aligned case, the disk-normal induction factor reduces to the standard axial induction factor $a = 1 - u_d / u_\infty$. 
The thrust force written in terms of the rotor-normal induction factor is then
\begin{equation}    
    \vec{F}_T = -\frac{1}{2} \rho C_T^{\prime} A_d ( 1 - a_n)^2 \cos^2(\gamma) u_\infty^2 \left[ \cos{(\gamma)} \hat{\imath} + \sin{(\gamma)} \hat{\jmath} \right].
    \label{eq:ft_full}
\end{equation}
The power for an actuator disk modeled wind turbine is computed as $P = -\vec{F}_T \cdot \vec{u}_d$.

The numerical ADM implementation follows the regularization methodology introduced by Calaf {\it et al.} \cite{calaf2010large} and further developed by Shapiro {\it et al.} \cite{shapiro2019filtered}. 
The porous disk thrust force $\vec{f}(\vec{x})=\vec{F}_T \mathcal{R}(\vec{x})$ is implemented in the domain ($\vec{x}$) through an indicator function $\mathcal{R}(\vec{x})$.
The indicator function $\mathcal{R}(\vec{x})$ is $\mathcal{R}(\vec{x}) = \mathcal{R}_1(x) \mathcal{R}_2(y, z)$ where
\begin{gather}
    \label{eq:R1_func}
    \mathcal{R}_1(x) = \frac{1}{2s} \left[
    \mathrm{erf}\left(\frac{\sqrt{6}}{\Delta}
    \left(x+\frac s2\right)\right) -
    \mathrm{erf}\left(\frac{\sqrt{6}}{\Delta}
    \left(x-\frac s2\right)\right)\right], 
    \\
    \label{eq:R2_func}
    \mathcal{R}_2(y, z) = \frac{4}{\pi D^2}\frac{6}{\pi\Delta^2} 
    \iint H\left(D/2 - \sqrt{y'^2+z'^2}\right) 
    \exp{\left(-6 \frac{(y-y')^2+(z-z')^2}{\Delta^2}\right)}\,\mathrm{d}y'\,\mathrm{d}z', 
\end{gather}
where $H(x)$ is the Heaviside function, $\mathrm{erf}(x)$ is the error function, $s=3\Delta x/2$ is the ADM disk thickness, and $\Delta$ is the ADM filter width. 
The disk velocity $\vec{u}_d$, used in the thrust force calculation Eq.~\eqref{eq:ft_full_les_code}, is calculated using the indicator function, $\vec{u}_d = M\iiint \mathcal{R}(\vec{x}) \vec{u}(\vec{x}) \,\mathrm{d}^3\vec{x}$, where $\vec{u}(\vec{x})$ is the filtered velocity in the LES domain and $M$ is a correction factor that depends on the filter width $\Delta$ \cite{shapiro2019filtered}.
Small values of filter width $\Delta$ tend towards the theoretical actuator disk representation (a true actuator disk represents the disk as infinitesimally thin with a discontinuity in forcing at radial position $r=R$), but lead to numerical oscillations in LES.
Given the nature of spatially distributing the thrust force with a larger value of $\Delta$, numerical implementations of the ADM typically underestimate the induction and therefore overestimate power production \cite{shapiro2019filtered, munters2017optimal}. 
The correction factor $M=\left( 1 + (C_T' \Delta/(2\sqrt{3\pi}D)\right)^{-1}$ derived by Shapiro {\it et al.} \cite{shapiro2019filtered} is used to correct this error by ensuring that the filtered ADM in Eq.~\eqref{eq:R1_func} sheds the same amount of vorticity as the infinitesimally thin disk, depending on $C_T^\prime$ and the ADM filter width.
Here, we use $\Delta/D = 3h/(2D)$ where $h = (\Delta x^2 + \Delta y^2 + \Delta z^2)^{1/2}$ is the effective grid spacing.
The qualitative and quantitative conclusions of this paper are not affected by this choice, as shown in {\bf Comparison between different actuator disk model regularization methods in LES} in the {\bf Supplementary Information}, provided that the correction factor $M$ derived by Shapiro {\it et al.} \cite{shapiro2019filtered} is used for larger $\Delta$. 
For small $\Delta$ values (e.g. $\Delta/D=0.032$), numerical (grid-to-grid) oscillations contaminate the wake pressure measurements.
Therefore, we have selected $\Delta/D = 3h/(2D)$ with $M$ given above.
In summary, sensitivity experiments are performed for different numerical ADM implementations in {\bf Supplementary Information}, and the results demonstrate that the qualitative conclusions of this study do not depend on the numerical implementation of the ADM.

\subsection*{Streamtube analysis numerical setup}
We consider a three dimensional streamtube analysis.
While a yaw aligned actuator disk in uniform inflow presents an axisymmetric streamtube, yaw misalignment results in wake curling \cite{howland2016wake} and three dimensional variations that motivate a three dimensional streamtube analysis.
A theoretical actuator disk model has uniform thrust force for all positions $r<R$ within the rotor, where $r$ is a radial position defined relative to the disk center and $R$ is the radius of the disk.
Numerical implementations of actuator disk models for computational fluid dynamics use regularization methodologies to avoid sharp discontinuities in the wind turbine body force \cite{heck2023modelling, calaf2010large, shapiro2019filtered}, as discussed in {\bf Large eddy simulation numerical setup}, which results in some variation in the thrust force towards the outer extent of the disk radius.
The degree to which there is variation in the thrust force for $r<R$ depends on the numerical implementation of the regularization (i.e. filter length $\Delta$ in the present implementation).
Following previous studies \cite{heck2023modelling, shapiro2018modelling}, to focus the streamtube analysis on the portion of the wake over which the thrust force is constant, the streamtube seed points are defined at the actuator disk with an initial radius of $R_s<R$, where $R_s/R=0.7$.
Numerical tests (not shown) demonstrate a small quantitative sensitivity between $0.5<R_s/R<0.9$, but the qualitative results are insensitive to $R_s$.

While flow quantities within the streamtube depend on $x$, a core assumption in near-wake models is that there is a particular $x$ location (or a range of $x$ locations), that is characteristic of the inviscid near-wake (potential core), such that near-wake flow quantities can be described with a single value per variable, rather than a one-dimensional field variable depending on $x$.
The fidelity of this assumption will be investigated in {\bf Streamtube analysis and budgets} in {\bf Supplementary Information}.
Here, we describe the methodology define these individual flow quantities.
The near-wake streamwise velocity $u_4$ is taken as the minimum value on the interval $1<x/D<2$, which is equivalent to the maximum streamwise velocity deficit $u_\infty - u_4$.
The wake pressure $p_4$ is calculated at the $x$ position where $u_4$ is minimum.
Following Shapiro {\it et al.}~\cite{shapiro2018modelling}, the lateral velocity $v_4$ is taken as its maximum, which is closer to the actuator disk, approximately at $x/D=0.5$ (see discussion by Shapiro {\it et al.}~\cite{shapiro2018modelling}).




\clearpage

\bibliography{scibib}

\begin{thebibliography}{10}

\bibitem{irena2022world}
IRENA, World energy transitions outlook 2022: 1.5° c pathway (2022).

\bibitem{iea2021net}
IEA., {\it Int. Energ. Agency\/} {\bf 224} (2021).

\bibitem{larson2020netzero}
E.~Larson, {\it et~al.\/}, Net-zero america: potential pathways,
  infrastructure, and impacts. (2020).

\bibitem{veers2019grand}
P.~Veers, {\it et~al.\/}, {\it Science\/} {\bf 366}, eaau2027 (2019).

\bibitem{meyers2022wind}
J.~Meyers, {\it et~al.\/}, {\it Wind Energy Science Discussions\/} {\bf 2022},
  1 (2022).

\bibitem{mai2017value}
T.~T. Mai, E.~J. Lantz, M.~Mowers, R.~Wiser, The value of wind technology
  innovation: Implications for the us power system, wind industry, electricity
  consumers, and environment, {\it Tech. rep.\/}, National Renewable Energy
  Lab.(NREL), Golden, CO (United States) (2017).

\bibitem{van2007lanchester}
G.~A. Van~Kuik, {\it Wind Energy: An International Journal for Progress and
  Applications in Wind Power Conversion Technology\/} {\bf 10}, 289 (2007).

\bibitem{glauert1926empirical}
H.~Glauert, The analysis of experimental results in the windmill brake and
  vortex ring states of an airscrew, {\it Tech. Rep. 1026\/}, ARCR R\&M (1926).

\bibitem{burton2011wind}
T.~Burton, N.~Jenkins, D.~Sharpe, E.~Bossanyi, {\it Wind energy handbook\/}
  (John Wiley \& Sons, 2011).

\bibitem{sorensen1998analysis}
J.~S{\o}rensen, W.~Shen, X.~Munduate, {\it Wind Energy: An International
  Journal for Progress and Applications in Wind Power Conversion Technology\/}
  {\bf 1}, 73 (1998).

\bibitem{martinez2022numerical}
L.~A. Mart{\'\i}nez-Tossas, {\it et~al.\/}, {\it Wind Energy\/} {\bf 25}, 605
  (2022).

\bibitem{moriarty2005aerodyn}
P.~J. Moriarty, A.~C. Hansen, Aerodyn theory manual, {\it Tech. rep.\/},
  National Renewable Energy Lab., Golden, CO (US) (2005).

\bibitem{madsen2020implementation}
H.~A. Madsen, T.~J. Larsen, G.~R. Pirrung, A.~Li, F.~Zahle, {\it Wind Energy
  Science\/} {\bf 5}, 1 (2020).

\bibitem{rankine1865}
{Rankine W.J.}, {Trans. Inst. Naval Arch.} (1865).

\bibitem{froude1878}
{Froude W.}, {Trans. Inst. Naval Arch.} (1878).

\bibitem{froude1889}
{Froude R.E.}, {Trans. Inst. Naval Arch.} (1889).

\bibitem{carlton2018marine}
J.~Carlton, {\it Marine propellers and propulsion\/} (Butterworth-Heinemann,
  2018).

\bibitem{glauert1926general}
H.~Glauert, {\it et~al.\/}, {\it A general theory of the autogyro\/}, vol. 1111
  (HM Stationery Office, 1926).

\bibitem{glauert1935airplane}
H.~Glauert, {\it Aerodynamic theory\/} (Springer, 1935), pp. 169--360.

\bibitem{johnson2012helicopter}
W.~Johnson, {\it Helicopter theory\/} (Courier Corporation, 2012).

\bibitem{amir2008modeling}
M.~Y. Amir, V.~Abbass, {\it 2008 International Conference on Smart
  Manufacturing Application\/} (IEEE, 2008), pp. 100--105.

\bibitem{pratumnopharat2011validation}
P.~Pratumnopharat, P.~S. Leung, {\it Renewable energy\/} {\bf 36}, 3222 (2011).

\bibitem{ross2020experimental}
H.~Ross, B.~Polagye, {\it Renewable Energy\/} {\bf 152}, 1328 (2020).

\bibitem{stoddard1978momentum}
F.~Stoddard, {\it Wind Technology Journal\/} {\bf 1}, 3 (1978).

\bibitem{wilson1974applied}
R.~E. Wilson, P.~Lissaman, {\it Oregon State University\/}  (1974).

\bibitem{shapiro2018modelling}
C.~R. Shapiro, D.~F. Gayme, C.~Meneveau, {\it J. Fluid Mech.\/} {\bf 841}
  (2018).

\bibitem{fleming2014field}
P.~Fleming, {\it et~al.\/}, {\it Journal of Physics: Conference Series\/} (IOP
  Publishing, 2014), vol. 524.

\bibitem{johlas2021floating}
H.~M. Johlas, L.~A. Mart{\'\i}nez-Tossas, M.~J. Churchfield, M.~A. Lackner,
  D.~P. Schmidt, {\it Wind Energy\/} {\bf 24}, 901 (2021).

\bibitem{doeoffshorewind}
D.~of~Energy, Floating offshore wind shot: Unlocking the power of floating
  offshore wind energy,
  \url{https://www.energy.gov/sites/default/files/2022-09/floating-offshore-wind-shot-fact-sheet.pdf}
  (2022).

\bibitem{dahlberg2005research}
J.~Dahlberg, B.~Montgomerie, {\it Swedish Defense Research Agency, FOI, Kista,
  Sweden, Report No. FOI\/} pp. 02--17 (2005).

\bibitem{liew2020analytical}
J.~Y. Liew, A.~M. Urb{\'a}n, S.~J. Andersen, {\it Wind Energy Science\/} {\bf
  5}, 427 (2020).

\bibitem{howland2020influencefield}
M.~F. Howland, {\it et~al.\/}, {\it Journal of Renewable and Sustainable
  Energy\/} {\bf 12}, 063307 (2020).

\bibitem{barthelmie2007modelling}
R.~J. Barthelmie, {\it et~al.\/}, {\it Wind Energy\/} {\bf 10}, 517 (2007).

\bibitem{barthelmie2010evaluation}
R.~J. Barthelmie, L.~Jensen, {\it Wind Energy\/} {\bf 13}, 573 (2010).

\bibitem{fleming2019initial}
P.~Fleming, {\it et~al.\/}, {\it Wind Energy Science\/} {\bf 4} (2019).

\bibitem{howland2022collective}
M.~F. Howland, {\it et~al.\/}, {\it Nature Energy\/} {\bf 7}, 818 (2022).

\bibitem{roshko1955wake}
A.~Roshko, {\it Journal of the aeronautical sciences\/} {\bf 22}, 124 (1955).

\bibitem{betz1920}
{Betz A.}, {Das Maximum der theoretisch m\"{o}glichen Ausn\"{u}tzung des Windes
  durch Windmotoren} (1920).

\bibitem{eggleston1987wind}
D.~M. Eggleston, F.~Stoddard, {\it Wind turbine engineering design\/} (Van
  Nostrand Reinhold Co. Inc., New York, NY, 1987).

\bibitem{buhl2005new}
M.~L. Buhl~Jr, New empirical relationship between thrust coefficient and
  induction factor for the turbulent windmill state, {\it Tech. rep.\/},
  National Renewable Energy Lab.(NREL), Golden, CO (United States) (2005).

\bibitem{bay2020floris}
C.~Bay, {\it et~al.\/}, Floris: A brief tutorial, {\it Tech. rep.\/}, National
  Renewable Energy Lab (NREL), Golden, CO (United States) (2020).

\bibitem{jonkman2005fast}
J.~M. Jonkman, M.~L. Buhl~Jr, {\it et~al.\/}, {\it Golden, CO: National
  Renewable Energy Laboratory\/} {\bf 365}, 366 (2005).

\bibitem{gebraad2016wind}
P.~Gebraad, {\it et~al.\/}, {\it Wind Energy\/} {\bf 19}, 95 (2016).

\bibitem{milne1973theoretical}
L.~M. Milne-Thomson, {\it Theoretical Aerodynamics\/} (Courier Corporation,
  1973).

\bibitem{heck2023modelling}
K.~S. Heck, H.~M. Johlas, M.~F. Howland, {\it Journal of Fluid Mechanics\/}
  {\bf 959}, A9 (2023).

\bibitem{steiros2018drag}
K.~Steiros, M.~Hultmark, {\it Journal of Fluid Mechanics\/} {\bf 853}, R3
  (2018).

\bibitem{madsen2023linear}
H.~A. Madsen, {\it Wind Energy Science Discussions\/} {\bf 2023}, 1 (2023).

\bibitem{durand1935aerodynamic}
W.~Durand, Aerodynamic theory: General aerodynamic theory: Perfect fluids/[by]
  th. von k{\'a}rm{\'a}n; jm burgers (1935).

\bibitem{bastankhah2016experimental}
M.~Bastankhah, F.~Port{\'e}-Agel, {\it J. Fluid Mech.\/} {\bf 806}, 506 (2016).

\bibitem{lee2003turbulent}
J.~H.-w. Lee, V.~H. Chu, {\it Turbulent jets and plumes: a Lagrangian
  approach\/}, vol.~1 (Springer Science \& Business Media, 2003).

\bibitem{calaf2010large}
M.~Calaf, C.~Meneveau, J.~Meyers, {\it Phys. Fluids\/} {\bf 22}, 015110 (2010).

\bibitem{taylor1944air}
G.~Taylor, {\it Aeronautical Research Council, Reports and Memoranda\/} {\bf
  2236}, 159 (1944).

\bibitem{koo1973fluid}
J.-K. Koo, D.~F. James, {\it Journal of Fluid Mechanics\/} {\bf 60}, 513
  (1973).

\bibitem{o2006source}
F.~O’Neill, {\it Ocean engineering\/} {\bf 33}, 1884 (2006).

\bibitem{larsen20072}
T.~J. Larsen, A.~M. Hansen, {\it Ris{\o} National Laboratory\/}  (2007).

\bibitem{pedersen2019dtuwindenergy}
M.~M. Pedersen, P.~van~der Laan, M.~Friis-M{\o}ller, J.~Rinker, P.-E.
  R{\'e}thor{\'e}, {\it Zenodo [code]\/} {\bf 10} (2019).

\bibitem{veers2023grand}
P.~Veers, {\it et~al.\/}, {\it Wind Energy Science\/} {\bf 8}, 1071 (2023).

\bibitem{liew_2024_10524067}
J.~Liew, K.~S.~Heck, M.~F. Howland, {Howland-Lab/Unified-Momentum-Model}
  (2024). Available at: \url{https://doi.org/10.5281/zenodo.10524066}.

\bibitem{ghate2018padeops}
A.~Ghate, A.~Subramaniam, M.~F. Howland, Padéops, GitHub [code] (2018).
  Available at: \url{https://github.com/Howland-Lab/PadeOps}.

\bibitem{ghate2017subfilter}
A.~S. Ghate, S.~K. Lele, {\it J. Fluid Mech.\/} {\bf 819}, 494 (2017).

\bibitem{howland2020influence}
M.~F. Howland, A.~S. Ghate, S.~K. Lele, {\it J. Fluid Mech.\/} {\bf 883}
  (2020).

\bibitem{nagarajan2003robust}
S.~Nagarajan, S.~K. Lele, J.~H. Ferziger, {\it J Comput Phys\/} {\bf 191}, 392
  (2003).

\bibitem{gottlieb2011strong}
S.~Gottlieb, D.~I. Ketcheson, C.-W. Shu, {\it Strong stability preserving
  Runge-Kutta and multistep time discretizations\/} (World Scientific, 2011).

\bibitem{nicoud2011using}
F.~Nicoud, H.~B. Toda, O.~Cabrit, S.~Bose, J.~Lee, {\it Phys. Fluids\/} {\bf
  23}, 085106 (2011).

\bibitem{nordstrom1999fringe}
J.~Nordstr{\"o}m, N.~Nordin, D.~Henningson, {\it SIAM Journal on Scientific
  Computing\/} {\bf 20}, 1365 (1999).

\bibitem{shapiro2019filtered}
C.~R. Shapiro, D.~F. Gayme, C.~Meneveau, {\it Wind Energy\/} {\bf 22}, 1414
  (2019).

\bibitem{munters2017optimal}
W.~Munters, J.~Meyers, {\it Philosophical Transactions of the Royal Society A:
  Mathematical, Physical and Engineering Sciences\/} {\bf 375}, 20160100
  (2017).

\bibitem{howland2016wake}
M.~F. Howland, J.~Bossuyt, L.~A. Mart{\'\i}nez-Tossas, J.~Meyers, C.~Meneveau,
  {\it J. Renew. Sustain. Energy\/} {\bf 8}, 043301 (2016).

\end{thebibliography}

\bibliographystyle{Science}

\clearpage


\end{document}